\documentclass[pra,twocolumn,showpacs,showkeys]{revtex4}
\usepackage{amsmath}
\usepackage{amsfonts}
\usepackage{graphicx}

\begin{document}

\title{Two-body and three-body substructures served as building blocks in
small spin-3 condensates}
\author{C. G. Bao}
\thanks{Corresponding author: stsbcg@mail.sysu.edu.cn}
\affiliation{Center of Theoretical Nuclear Physics, National Laboratory of Heavy Ion
Accelerator, Lanzhou, 730000, People's Republic of China}
\affiliation{State Key Laboratory of Optoelectronic Materials and Technologies, School of
Physics and Engineering, Sun Yat-Sen University, Guangzhou, 510275, People's
Republic of China}
\date{\today }

\begin{abstract}
It was found that stable few-body spin-structures, pairs and triplexes, may
exist as basic constituents in small spin-3 condensates, and they will play
the role as building blocks when the parameters of interaction are
appropriate. Specific method is designed to find out these constituents.
\end{abstract}

\maketitle

\section{Introduction}

When the structure of a few-body system is very stable, it may
become a basic constituent of many-body systems. For an
example, the structures of light nuclei can be explained based
on the cluster model, where the $\alpha$-particle is a building
block.\cite{r_EWS,r_NDC} Another famous example is the Cooper
pair in condensed matter.\cite{r_LNC,r_AMK} This pair is
responsible for the superconductivity. For Bose-Einstein
condensates of spin-1 and spin-2 atoms, basic few-body
structures have already been proposed by theorists. For spin-1
condensates, the interaction can be written as
\begin{eqnarray}
 V_{ij}
  =  \delta(\textbf{\textit{r}}_i-\textbf{\textit{r}}_j)
     \sum_S
     g_S
     \mathcal{P}_S,
 \label{e01_Vij}
\end{eqnarray}
where $S$ is the combined spin of $i$ and $j$ and has two
choices 0 and 2. $\mathcal{P}_S$ is the projector of the
$S$-channel, $g_S$ is the strength proportional to the $s$-wave
scattering length of the $S$-channel. When $g_2-g_0$ is
positive, the ground state will have total spin $F=0$ (1) when
the particle number $N$ is even
(odd).\cite{r_CKL,r_CGB,r_PVI,r_JKA} The ground state wave
function $\Psi_g$ is proportional to the pair-state
$\tilde{P}_N[(\varsigma \varsigma)_0]^{N/2}$ or
$\tilde{P}_N\varsigma[(\varsigma\varsigma)_0]^{(N-1)/2}$, where
$\varsigma$ denotes the spin-state of a spin-1 atom, a pair of
them are coupled to zero, $\tilde{P}_N$ is the symmetrizer (a
summation over the $N!$ permutation terms). For higher states,
say, the one with $F=2$ is proportional to
$\tilde{P}_N[(\varsigma\varsigma)_2(\varsigma\varsigma)_0]^{(N-2)/2}$.
Thus, the singlet pair $(\varsigma\varsigma)_0$ appears as a
common building block. These pairs, together with a few other
substructures, constitute the low-lying states.

For spin-2 condensates, the interaction can also be written as
Eq.~(\ref{e01_Vij}), but with $S=0$, $2$, and $4$. A detailed
classification of the spin-states based on the seniority and
$F$ was given in the Ref.~\cite{r_PVI}. When $N$ is even and
$\frac{7}{10}(g_0-g_4)<(g_2-g_4)<-\frac{7(N+3)}{10(N-2)}(g_0-g_4)$,
$\Psi_g$ is also dominated by the singlet pairs, and is
proportional to the pair-state
$\tilde{P}_N[(\eta\eta)_0]^{N/2}$, where $\eta$ denotes the
spin-state of a spin-2 atom. When $N$ is a multiple of 3, and
$(g_2-g_4)$ is negative and smaller than
$\frac{7}{10}(g_0-g_4)$, $\Psi_g$ is nearly proportional to the
triplex-state $\tilde{P}_N[((\eta\eta)_2\eta)_0]^{N/3}$.
Obviously, the triplex $((\eta\eta)_2\eta)_0$ acts as a
building block. Furthermore, when the parameters of interaction
fall inside the indicated domain, the low-lying states are also
dominated by these building blocks accompanied by a few other
substructures.

\begin{figure}[htbp]
 \centering
 \resizebox{0.55\columnwidth}{!}{\includegraphics{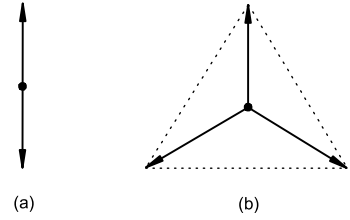}}
 \caption{Intuitive diagram of basic constituents of
condensates: the pair (a) and triplex (b).}
 \label{fig1}
\end{figure}

It is evident that the existence of building blocks originates
from the special feature of interaction. The pairs
$(\varsigma\varsigma)_S$ and $(\eta\eta)_S$ may appear as
building blocks whenever $g_S$ is sufficiently negative.
However, for spin-$f$ atoms with $f$ even, instead of the
pairs, a more favorable substructure is a triplex. Let
$\vartheta$ denote the spin-state of a spin-$f$ particle. For
the triplex-state
$\tilde{P}_N((\vartheta\vartheta)_f\vartheta)_{\lambda}$, the
symmetrizer $\tilde{P}_N$ is not necessary when $\lambda=0$
because $((\vartheta\vartheta)_f\vartheta)_0$ itself is
symmetric (i.e., $((\vartheta(i)\vartheta(j))_f\vartheta(k))_0
=((\vartheta(j)\vartheta(k))_f\vartheta(i))_0$ as can be
verified by recoupling the spins). It implies that every two
spins are coupled to $f$. Thereby the binding would be
maximized if $g_f$ is sufficiently negative. Intuitively
speaking, the three spins in the $\lambda=0$ triplex will be
coplanar and form a regular triangle. This is shown in
Fig.~\ref{fig1}b, where the angle between every two spins is
120$^{\circ}$ so that they are coupled to $f$. Therefore, for
spin-$f$ condensates, the $\lambda=0$ triplex might serve as
building blocks when $g_f$ is sufficiently negative (note that
the $\lambda=0$ triplex is prohibited when $f$ is odd).

The ground state structure of spin-3 condensates has already
been studied based on the MFT.\cite{r_RBD,r_LSA,r_HMA} In this
theory the spin structure is described by the spinors. They
will have the polar phase (corresponding to the pair-state)
when $g_0$ is sufficiently negative, and have the cyclic phase
(corresponding to the triplex-state) when $g_0$ becomes
positive. If a magnetic field is applied, a number of phases
will emerge. Instead of using the spinors, it is an attempt of
this paper to describe the spin structures based on the basic
constituents. From the experience of spin-1 and -2 condensates,
it is expected that pairs and triplexes might also appear in
spin-3 condensates when the parameters of interaction are
appropriate. We are going to search for these basic
constituents. Due to the difficulty in calculation, only small
condensates ($N$ is small) are concerned. We believe that the
knowledge from small systems will help us to understand better
the larger systems. Due to the prohibition of the $\lambda=0$
triplex, only the $\lambda\neq 0$ triplexes could emerge in
spin-3 systems. Since each $\lambda$-triplex has $2\lambda+1$
magnetic components, additional complexity will arise as shown
below.

\section{Hamiltonian, eigenstates and particle correlation}

Let $N$ spin-3 atoms be confined by an isotropic and parabolic
trap with frequency $\omega$. The interaction between particle
$i$ and $j$ is $V_{ij}
=\delta(\textbf{\textit{r}}_i-\textbf{\textit{r}}_j)\sum_S
g_S\mathcal{P}_S+V_{\mathrm{dd}}$, where $V_{\mathrm{dd}}$ is
the dipole-dipole interaction. Let the wave function for the
relative motion of $i$ and $j$ be $\psi_{lSJ}$, where $l$ is
the relative orbital angular momentum, and $l$ the $S$ are
coupled to $J$. Then, for $l=0$, the matrix element
$\langle\psi_{l'S'J}|V_{\mathrm{dd}}|\psi_{0SJ}\rangle$ is
non-zero only if $l'=2$. It implies that $V_{\mathrm{dd}}$
would play its role only if a $d$-wave spatial excitation is
accompanied. It is assumed that $\omega$ is so large that
$\langle\psi_{2S'J}|V_{\mathrm{dd}}|\psi_{0SJ}\rangle<<2\hbar\omega$.
In this case the effect of $V_{\mathrm{dd}}$ is suppressed so
that it can be neglected. Note that, in general,
$V_{\mathrm{dd}}$ would be important in spin evolution where
higher partial waves emerge. However, this is not the case of
equilibrium state in strong isotropic trap at very low
temperature.\cite{r_LSA,r_HMA,r_Z} Furthermore, we consider a
small condensate so that the size of the condensate is smaller
than the spin healing length. In this case the single spatial
mode approximation (SMA) is reasonable and is therefore
adopted.\cite{r_MSC}

Let the common spatial wave function be
$\phi(\textbf{\textit{r}})$. Under the SMA, after an
integration over the spatial degrees of freedom, we arrive at a
model Hamiltonian
\begin{eqnarray}
 H_{\mathrm{mod}}
  =  \sum_{i<j}
     V'_{ij},
 \label{e02_H}
\end{eqnarray}
where $V'_{ij}=\sum_S G_S\mathcal{P}_S$,
$G_S=g_S\int|\phi(\textbf{\textit{r})}|^4\mathrm{d}\textbf{\textit{r}}$.

For diagonalizing $H_{\mathrm{mod}}$, the set of normalized and
symmetrized Fock-states $|\alpha\rangle
=|N_3^{\alpha},N_2^{\alpha},N_1^{\alpha},N_0^{\alpha},N_{-1}^{\alpha},N_{-2}^{\alpha},N_{-3}^{\alpha}\rangle$
are used as basis functions, where $N_{\mu}^{\alpha}$ is the
number of particles in the $\mu$-spin-component,
$\sum_{\mu}N_{\mu}^{\alpha}=N$ and $\sum_{\mu}\mu
N_{\mu}^{\alpha}=M$, where $M$ is the magnetization. Since
$|\alpha\rangle$ as a whole form a complete set, once the
matrix elements $\langle\alpha'|H_{\mathrm{mod}}|\alpha\rangle$
have been calculated, exact eigenenergies and eigenstates of
$H_{\mathrm{mod}}$ can be obtained via the diagonalization.
Obviously, both the total spin $F$ and its $Z$-component $M$
are conserved when $V_{\mathrm{dd}}$ is neglected.

Let the $i$-th eigenstate be denoted as $\psi_i$ with total
spin $F(i)$ and magnetization $M(i)$. The state can be expanded
as $\psi_i=\sum_{\alpha}c_{\alpha}|\alpha\rangle$. Since one
can extract a particle from a Fock-state, one can also extract
a particle from $\psi_i$ via the expansion as
\begin{eqnarray}
 \psi_i
 &\equiv&
     \sum_{\mu}
     \chi_{\mu}(1)
     \psi_{\mu}^i,  \nonumber \\
 \psi_{\mu}^i
 &=& \sum_{\alpha}
     c_{\alpha}
     \sqrt{\frac{N_{\mu}^{\alpha }}{N}}
     |\cdots,N_{\mu}^{\alpha}-1,\cdots\rangle,
 \label{e03_psii}
\end{eqnarray}
where $\chi_{\mu}$ is the spin-state of a $f=3$ particle in
component $\mu$ (from $-3$ to $3$). With Eq.~(\ref{e03_psii}),
we know that the probability of a particle in $\mu$ is just
\begin{equation}
 P_{\mu }^i
  \equiv
      \langle
      \psi_{\mu}^i |
      \psi_{\mu}^i
      \rangle.
 \label{e04_pui}
\end{equation}
They fulfill $\sum_{\mu}P_{\mu}^i=1$. $P_{\mu}^i$ is called the
1-body probability, and $N P_{\mu}^i$ is just the average
population of the $\mu$ component.

If one more particle is further extracted, in a similar way, we have
\begin{eqnarray}
 \psi_i
  =  \sum_{\mu,\nu}
     \chi_{\mu}(1)
     \chi_{\nu}(2)
     \varphi_{\mu\nu}^i.
 \label{e05_psii}
\end{eqnarray}
We define
\begin{eqnarray}
 P_{\mu\nu}^i
  \equiv
     \langle
     \varphi_{\mu\nu}^i |
     \varphi_{\mu\nu}^i
     \rangle.
 \label{e06_Puvi}
\end{eqnarray}
They fulfill $P_{\mu\nu}^i=P_{\nu\mu}^i$, $\sum_{\mu\nu}P_{\mu\nu}^i=1$, and
$\sum_{\nu}P_{\mu\nu}^i=P_{\mu}^i$. When two particles are observed
simultaneously, obviously, the probability of one in $\mu$ and the other one
in $\nu$ is $P_{\mu\nu}^i+P_{\nu\mu}^i$ (if $\mu\neq\nu$) or $P_{\mu\mu}^i$
(if $\mu=\nu$). $P_{\mu\nu}^i$ is called the correlative probability of
spin-components.

Let Eq.~(\ref{e05_psii}) be rewritten as
\begin{equation}
 \psi_i
  =  \sum_{S,m_S}
     (\chi(1)\chi(2))_{S,m_S}
     \sum_{\mu}
     C_{3,\mu,3,m_S-\mu}^{S,m_S}
     \varphi_{\mu,m_S-\mu}^i.
 \label{e07_psii}
\end{equation}
Then, the probability of a pair of particles coupled to $S$ and
$m_S$ is
\begin{eqnarray}
 P_{S,m_S}^i
 &=& \sum_{\mu',\mu}
     C_{3,\mu',3,m_S-\mu'}^{S,m_S}
     C_{3,\mu,3,m_S-\mu}^{S,m_S}  \nonumber \\
  && \langle
     \varphi_{\mu',m_S-\mu'}^i |
     \varphi_{\mu,m_S-\mu}^i\rangle.
 \label{e08_Psms}
\end{eqnarray}
One can prove from symmetry that $P_{S,m_S}^i=P_{S,-m_S}^i$
when $M(i)=0$, and the $2S+1$ members $P_{S,m_S}^i$ are equal
to each other when $F(i)=0$. Furthermore, we define
\begin{equation}
 \mathfrak{P}_S^i
  =  \sum_{m_S}
     P_{S,m_S}^i,
 \label{e09_PSi}
\end{equation}
which is the probability that the spins of an arbitrary pair
are coupled to $S$. In general, $\mathfrak{P}_S^i$ can provide
information on the possible existence of two-body substructures
as shown below.

As an example, let the normalized pair-state be denoted as
\begin{equation}
 \Psi_{\mathrm{polar}}
  =  \gamma
     \tilde{P}_N
     [(\chi\chi)_0]^{N/2},
 \label{e10_Psipolar}
\end{equation}
where
$\gamma=[N!(\frac{2}{7})^{N/2}(N/2)!\frac{(N+5)!!}{5!!}]^{-1/2}$
is the constant for normalization (refer to Appendix I). Since
\begin{eqnarray}
 \tilde{P}_N[(\chi\chi)_0]^{N/2}
 &=& \sum_{S,m_S}
     (\chi(1)\chi(2))_{S,m_S}B_{S,m_S}  \nonumber \\
  && \tilde{P}_{N-2}
     (\chi\chi)_{S,m_S}
     [(\chi\chi)_0]^{(N-4)/2},
 \label{e11_PN}
\end{eqnarray}
where
\begin{equation}
 B_{S,m_S}
  =  N
     ( \delta_{S,0}
       \delta_{m_S,0}
      +(-1)^{m_S}
       2(N/2-1)/7.
 \label{e12_BSmS}
\end{equation}

By making use of the formulae in Appendix I, one can prove that
all the $P_{S,m_S}^i$ of the pair-state are equal to
$\frac{2(N-2)}{63(N-1)}$ with the only exception
$P_{0,0}^i=\frac{9(N+5)}{63(N-1)}$. From these data,
$\mathfrak{P}_S^i$ of the pair-state, denoted as
$\mathfrak{P}_S^{\mathrm{polar}}$, are listed in
Tab.~\ref{tab1}.
\begin{table}[tbh]
\caption{$\mathfrak{P}_S^{\mathrm{polar}}$, the probability
that a pair of particles in the polar state are coupled to
$S$.} \label{tab1}
\begin{ruledtabular}
 \begin{tabular}{lllll}
  $S$                                &  0                         &  2                          &  4                          & 6  \\
  \hline
  $\mathfrak{P}_S^{\mathrm{polar}}$  &  $\frac{9(N+5)}{63(N-1)}$  &  $\frac{10(N-2)}{63(N-1)}$  &  $\frac{18(N-2)}{63(N-1)}$  &  $\frac{26(N-2)}{63(N-1)}$ \\
 \end{tabular}
 \end{ruledtabular}
\end{table}

Note that, if the two particles are completely free, simply
from the geometry, we would have
$\mathfrak{P}_S^{\mathrm{free}}=(2S+1)/\sum_{S'}(2S'+1)$, where
$S'$ covers 0, 2, 4, and 6. Therefore,
$\mathfrak{P}_S^{\mathrm{free}}=0.036$, $0.179$, $0.321$, and
$0.464$ for $S=0$, $2$, $4$, and $6$, respectively. Whereas for
the pair-state, $\mathfrak{P}_S^{\mathrm{polar}}=0.143$,
$0.158$, $0.286$, and $0.413$ when $N\rightarrow\infty$. Thus
the ratios
$\mathfrak{P}_S^{\mathrm{polar}}/\mathfrak{P}_S^{\mathrm{free}}$
are $4.0$, $0.89$, $0.89$, and $0.89$, where the ratio with
$S=0$ is remarkably large. Thus the big ratio is a signal of
the dominance of the singlet pairs.

The spin-spin correlation is not able to be studied perfectly by the MFT.
The 1-body and correlative probabilities together will help us to understand
better the spin-structures. In the fields of atomic and nuclear physics, it
was found that the structures depend on $N$ sensitively (say, the property
of an even-even nucleus is quite different from its neighboring even-odd
nuclei). Since $N$ is assumed to be small in this paper, the $N$-dependence
of condensates is also studied in the follows.

\section{Populations of the spin-components of the ground states}

We shall first study the $^{52}$Cr atoms as a representative of
spin-3 species. These atoms have
$g_6=59.40meV\cdot\mathring{A}^3$, $g_4=0.5178 g_6$, and
$g_2=-0.0625 g_6$, while $g_0$ is unknown. In what follows
$G_6=g_6\int|\phi(\textbf{\textit{r}})|^4\mathrm{d}\textbf{\textit{r}}$
is considered as a unit of energy, and $g_0$ is variable. The
1-body probabilities $P_{\nu}^i$ are firstly studied.

When $N$ is even and $g_0\rightarrow-\infty$, it has been
proved that the ground state $\psi_1$ is exactly the pair-state
$\Psi_{\mathrm{polar}}$ written in
Eq.~(\ref{e10_Psipolar}).\cite{r_PVI,r_CGB2} Obviously, the
pair-state has $F=0$. It is a common feature that $P_{\nu}^i$
of all the $F=0$ states do not depend on $\nu$ due to the
isotropism. Therefore, all of them are equal to $1/(2f+1)=1/7$.

When $N$ is odd and $g_0\rightarrow-\infty$, $\psi_1$ is just
the odd pair-state written as
\begin{equation}
 \Psi_{\mathrm{odd-polar},\mu}
  =  \gamma'
     \tilde{P}_N
     \chi_{\mu}
     [(\chi\chi)_0]^{(N-1)/2},
 \label{e13_Psiodd}
\end{equation}
where
$\gamma'=[N!(\frac{2}{7})^{(N-1)/2}(\frac{N-1}{2})!\frac{(N+6)!!}{7!!}]^{-1/2}$
is the constant of normalization (refer to Appendix I).
Obviously, this state has $F=3$. The associated 1-body
probability $P_{\nu}^{\mathrm{odd-polar},\mu}$ has an
analytical form as given in Tab.~\ref{tab2} (the derivation is
referred to Appendix II). For the case with a large $N$, it is
shown in this table that $P_{\nu}^{\mathrm{odd-polar},\mu}$
depend on $N$ very weakly. When the spin of the unpaired
particle has $\mu=0$, we know from Tab.~\ref{tab2} that the
average population with $\nu=0$ is the largest and is nearly
third times as large as those with $\nu\neq 0$. When $\mu\neq
0$, the populations with $\nu=\mu$ and $-\mu$ are the largest
two and they are nearly two times as large as those with
$\nu\neq|\mu|$. On the other hand, it is recalled that
$P_{\nu}^{\mathrm{polar}}=1/7$ which is greatly different from
$P_{\nu}^{\mathrm{odd-polar},\mu}$. Thus, the populations of
spin-components of a pair-state will undergo a great change
when a particle with a given $\mu$ is added into the state.
This will happen even when $N$ is very large. Such special
even-odd dependence could be revealed by measuring the 1-body
probabilities if the polar-state can be prepared.
\begin{table}[tbh]
\caption{1-body probability $P_{\nu}^{\mathrm{odd-polar},\mu}$,
where $\mu$ is the spin-component of the single (unpaired)
particle in the odd polar state, and $\nu$ denotes the
component of the particle under observation.} \label{tab2}
\begin{ruledtabular}
 \begin{tabular}{cccccc}
  $\mu\neq 0$        &  $\mu\neq 0$          &  $\mu\neq 0$       &  &  $\mu=0$              &  $\mu=0$  \\
  $\nu=\mu$          &  $\nu=-\mu$           &  $\nu\neq|\mu|$    &  &  $\nu=0$              &  $\nu\neq 0$  \\
  \hline
  $\frac{2N+7}{9N}$  &  $\frac{2(N-1)}{9N}$  &  $\frac{N-1}{9N}$  &  &  $\frac{3(N+2)}{9N}$  &  $\frac{N-1}{9N}$  \\
 \end{tabular}
 \end{ruledtabular}
\end{table}

When $g_0$ increases from $-\infty$ but still negative,
$\psi_1$ with an even $N$ would be still more or less close to
$\Psi_{\mathrm{polar}}$. If the magnitudes of $g_2$, $g_4$, and
$g_6$ are small, then $\psi_1$ would be closer. However, for
realistic $^{52}$Cr, $g_4$ and $g_6$ are not small. To see how
large $\psi_1$ would deviate from the pair-state,
$\langle\Psi_{\mathrm{polar}}|\psi_1\rangle$ has been
calculated. If $g_0/g_6=-1$, $-2$, and $-4$, respectively,
$\langle\Psi_{\mathrm{polar}}|\psi_1\rangle=0.838$, $0.936$,
and $0.981$ when $N=12$. This set of values would become
$0.706$, $0.841$, and $0.937$ when $N=18$. Thus the deviation
is not small unless $g_0$ is very negative.

\begin{figure}[htbp]
 \centering
 \resizebox{0.95\columnwidth}{!}{\includegraphics{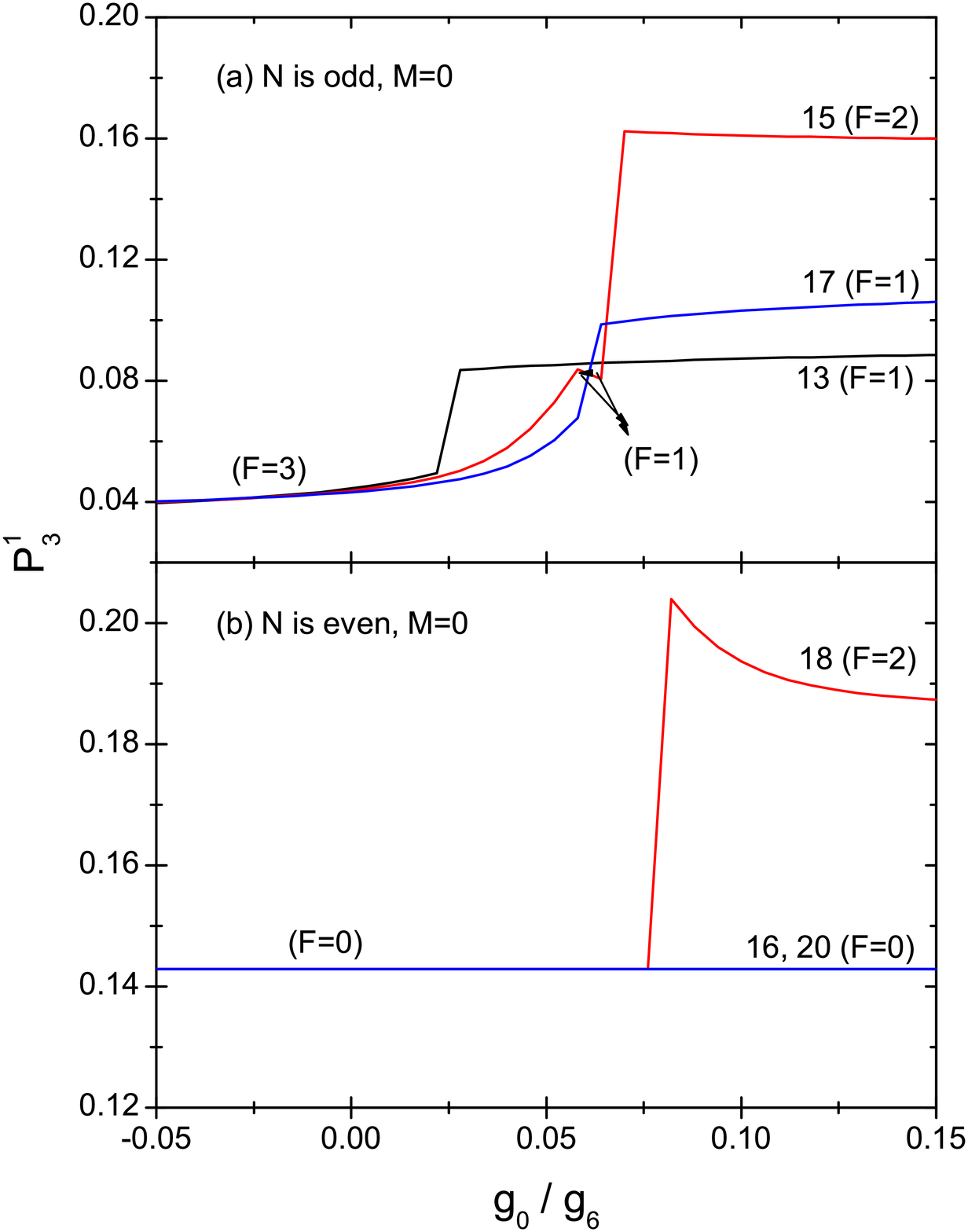}}
 \caption{(Color online) $P_3^1$ of the ground state $\psi_1$
with $M=0$ against $g_0/g_6$. $g_2$, $g_4$, and $g_6$ are given
at the experimental values of $^{52}$Cr. $N$ is marked by the
curves. The total spin $F$ of $\psi_1$ is given inside the
parentheses, it will jump whenever $g_0$ crosses a critical
point. Note that $N P_{\nu}^i$ is the average population of the
$\nu$-component of the $\psi_i$ state.}
 \label{fig2}
\end{figure}

For $^{52}$Cr, the 1-body probabilities $P_3^1$ of $\psi_1$
with $M=0$ against $g_0/g_6$ are plotted in Fig.~\ref{fig2}.
The curves at the left side of Fig.~\ref{fig2}b with an even
$N$ are horizontal lines. They have the same value $1/7$
implying that $\psi_1$ keeps $F=0$. The curves at the left side
of Fig.~\ref{fig2}a with an odd $N$ are close to each other
implying a weak dependence on $N$. They are flat and have their
values $\approx 0.038$ , which deviates explicitly from
$P_3^{\mathrm{odd-polar},0}\approx 0.103$ given in
Tab.~\ref{tab2}. It implies that the deviation between $\psi_1$
and $\Psi_{\mathrm{odd-polar},0}$ is not small when $g_0=-0.1$.

By comparing the right side of Fig.~\ref{fig2}a with that of
\ref{fig2}b, the even-odd dependence is clearly shown. In
addition, the curves with $N=15$ and $18$ are distinguished. It
implies that $N=3K$ ($K$ is an integer) is special. It is
recalled that, for spin-2 condensates with a sufficiently
negative $g_2$ , the ground state is formed by the triplex
$[(\eta\eta)_2\eta]_0$. \cite{r_MUE,r_PVI,r_CGB3} The finding
of the $3K$-dependence in spin-3 condensates is a hint that
3-body substructures might exist as well.

In Fig.~\ref{fig2} the domain of $g_0$ is roughly divided into
three regions. At the left side (region I) the curves depend on
$g_0$ mildly, and the singlet pairs play an important role. At
the right side (region III) the curves depend on $g_0$ also
mildly, where three-body substructures might exist. In between
(region II, roughly from $g_0/g_6=$ 0 to $0.1$) the curves vary
with $g_0$ very swiftly, implying a swift change in
spin-structure. In addition, critical point may appear in this
region (say, $g_0/g_6=0.079$ is a critical point when $N=18$
and $M=0$), where $P_3^1$ varies abruptly. Once $g_0$ crosses a
critical point $F$ will change suddenly implying a transition
of spin structure.

Incidentally, according to the MFT, there are also three
regions. The phases of the ground state in these regions are
named maximum polar(A), collinear polar(B), and biaxial
nematic(C) in the Ref.~\cite{r_RBD} when $g_0$ varies from
negative to positive, The associated spinors are
$(1,0,0,0,0,0,1)$, $(a,0,be^{i\delta},0,be^{i\delta},0,a)$, and
$(a,0,b,0,c,0,d)$. The critical point between A and B is
$g_0/g_6=0.079$. In the Ref.~\cite{r_LSA}, the three phases are
named the polar phase, the mixed phase, and the cyclic phase.
The associated spinors are $(\cos\theta,0,0,0,0,0,\sin\theta)$,
$(0,a,0,b,0,a,0)$ and $(a,0,b,0,b,0,a)$, respectively (the
third spinor may mix up with the second spinor in the mixed
phase). In this paper a different language is used so that the
structures can be understood via a different path.

\section{Correlative probabilities of the ground states}

The curves at the left side of Fig.~\ref{fig2}b are horizontal
until the critical point. However, the spin-structures are in
fact changing in this broad region. This example demonstrates
that the information provided by the 1-body probabilities is
not sufficient. Therefore, the correlative probabilities are
further studied. Firstly, for the pair-state, the probabilities
$P_{\mu\nu}^{\mathrm{polar}}$ are given in Tab.~\ref{tab3} (the
derivation is referred to Appendix III).
\begin{table}[tbh]
 \caption{$P_{\mu\nu}^{\mathrm{polar}}$, the correlative
probabilities of the polar state.}
 \label{tab3}
\begin{ruledtabular}
 \begin{tabular}{lllll}
  $(\mu,\nu)$                    &  $(0,0)$                   &  $(\mu\neq 0,-\mu)$      &  $(\mu\neq 0,\mu)$         &  otherwise  \\
  \hline
  $P_{\mu\nu}^{\mathrm{polar}}$  &  $\frac{3(N+1)}{63(N-1)}$  &  $\frac{2N+5}{63(N-1)}$  &  $\frac{2(N-2)}{63(N-1)}$  &  $\frac{N-2}{63(N-1)}$  \\
 \end{tabular}
 \end{ruledtabular}
\end{table}

It is shown that $P_{\mu\nu}^{\mathrm{polar}}$ are all nearly
independent on $N$ unless $N$ is small. The largest component
is $P_{0,0}^{\mathrm{polar}}$. The probabilities of being
spin-parallel and spin-anti-parallel, i.e.,
$P_{\mu,\mu}^{\mathrm{polar}}$ and
$P_{\mu,-\mu}^{\mathrm{polar}}$, are equal when
$N\rightarrow\infty$.

\begin{figure}[htbp]
 \centering
 \resizebox{0.95\columnwidth}{!}{\includegraphics{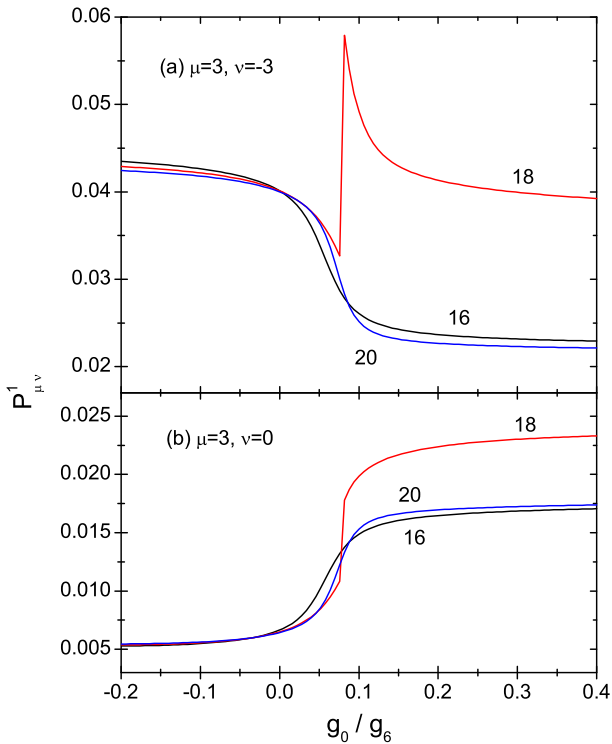}}
 \caption{(Color online) $P_{\mu\nu}^1$ of the ground state
$\psi_1$ of $^{52}$Cr against $g_0/g_6$. $N$ is even and is
marked by the side of each curve, and $M=0$.}
 \label{fig3}
\end{figure}

Examples of $P_{\mu\nu}^1$ of $\psi_1$ with an even $N$ are
given in Fig.~\ref{fig3}. The curves with $N=18$ are also
distinguished and jump up suddenly at the critical point. The
jump is accompanied by a change of the total spin $F$ from $0$
to 2 implying a transition. Since the curves vary rapidly in
the neighborhood of the critical point, strong adjustment in
structure happens right before and after the transition. The
strong adjustment in the neiborhood of the critical point is a
notable phenomenon. The curves with $N=16$ and 20 are similar
to each other. In particular, they keep their $F=0$ and
accordingly they do not have the sudden jump. The critical
point will appear whenever $N$ is a multiple of 3, and will
shift a little to the left when $N$ becomes larger (say, they
appear at $g_0/g_6=0.252$, $0.079$, and $0.075$, respectively,
when $N=12$, 18, and 24). The shift would be very small if
$\Delta N/N$ is small.

The curves of $P_{\mu\nu}^1$ with an odd $N$ are in general
very different from those with an even $N$. They also exhibit
the $3K$-dependence and contain the critical points in region
II. For an example, there are two critical points appearing at
$g_0/g_6=0.058$ and $0.067$ when $N=15$ and $M=0$. Accordingly,
$F$ jumps from 3 to 1, then to 2 when $g_0$ increases.

\section{Mixing of $S=0$ and 2 pairs}

Let us study $\mathfrak{P}_S^i$, the probabilities of the spins
of two particles coupled to $S$. $\mathfrak{P}_S^1$ of $\psi_1$
is close to $\mathfrak{P}_S^{\mathrm{polar}}$ when $N$ is even
and $g_0\rightarrow-\infty$. When $g_0$ is not so negative,
examples of $\mathfrak{P}_S^1/\mathfrak{P}_S^{\mathrm{free}}$
are listed in Tab.~\ref{tab4}. Where, $g_0/g_6=-0.2$ is in
region I, $0.07$ and $0.09$ are in region II and lying by the
left and right sides of the critical point (at $0.079$), $0.5$
and 1 are in region III.
\begin{table}[tbh]
 \caption{$\mathfrak{P}_S^1/\mathfrak{P}_S^{\mathrm{free}}$ of
the ground states of $^{52}$Cr with $N=18$, $M=0$, and $g_0$ is
given at five presumed values.}
 \label{tab4}
\begin{ruledtabular}
 \begin{tabular}{lllll}
  $\mathfrak{P}_S^1/\mathfrak{P}_S^{\mathrm{free}}$  &  $S=0$   &  $S=2$   &  $S=4$   &  $S=6$  \\
  \hline
  $g_0/g_6=-0.2$                                     &  $4.58$  &  $1.78$  &  $0.14$  &  $1.02$  \\
  $g_0/g_6=0.07$                                     &  $2.44$  &  $1.90$  &  $0.50$  &  $0.89$  \\
  $g_0/g_6=0.09$                                     &  $0.75$  &  $1.92$  &  $0.83$  &  $0.78$  \\
  $g_0/g_6=0.5$                                      &  $0.04$  &  $1.88$  &  $1.02$  &  $0.72$  \\
  $g_0/g_6=1$                                        &  $0.01$  &  $1.87$  &  $1.04$  &  $0.72$  \\
 \end{tabular}
 \end{ruledtabular}
\end{table}

We found that $\mathfrak{P}_0^1/\mathfrak{P}_0^{\mathrm{free}}$
is quite large when $g_0/g_6=-0.2$. It implies the preference
for the singlet pairs. However, the four ratios
$\mathfrak{P}_S^1/\mathfrak{P}_S^{\mathrm{free}}$ as a whole
deviate explicitly from
$\mathfrak{P}_S^{\mathrm{polar}}/\mathfrak{P}_S^{\mathrm{free}}$.
Therefore $\psi_1$ is quite different from the pair-state. When
$g_0$ increases further, the ratio with $S=0$ keeps on
decreasing. In particular, it decreases very rapidly when $g_0$
is passing through the critical point. On the other hand, the
probability of $S=2$ pair remains to be larger. In particular,
it becomes the largest when $g_0$ is larger than the critical
value. Therefore, the dominance of the singlet pairs is
gradually replaced by the $S=2$ pairs. Hence, for $N$ being
even, we define a set of basis functions formed by the two
kinds of pairs as
\begin{eqnarray}
 \Phi_j^{\mathrm{pairs}}
 &=& \beta_j
     \tilde{P}_N
     \{ [(\chi\chi)_0]^{K_p}
        [(\chi\chi)_{2,2}]^{K_2}  \nonumber \\
  &&    \cdots
        [(\chi\chi)_{2,-2}]^{K_{-2}} \},
 \label{e14_Phij}
\end{eqnarray}
where $j$ denotes the set $(K_p,K_2,\cdots,K_{-2})$ of
non-negative integers, and their sum $=N/2$. $\beta_j$ is for
the normalization. The space expanded by
$\Phi_j^{\mathrm{pairs}}$ is much smaller than that expanded by
the Fock-state. Say, if $N=18$ and $M=0$, the numbers of
Fock-states and $\Phi_j^{\mathrm{pairs}}$ are 3486 and 148,
respectively. The eigenstates can be approximately expanded as
\begin{equation}
 \psi_i
  \approx
     \sum_j
     b_j
     \Phi_j^{\mathrm{pairs}}
  \equiv
     \tilde{\psi}_i,
 \label{e15_psii}
\end{equation}
where $b_j$ can be obtained via a diagonalization of
$H_{\mathrm{mod}}$ in the much smaller space (Note that
$\Phi_j^{\mathrm{pairs}}$ are not exactly orthogonal to each
other).

\begin{figure}[htbp]
 \centering
 \resizebox{0.95\columnwidth}{!}{\includegraphics{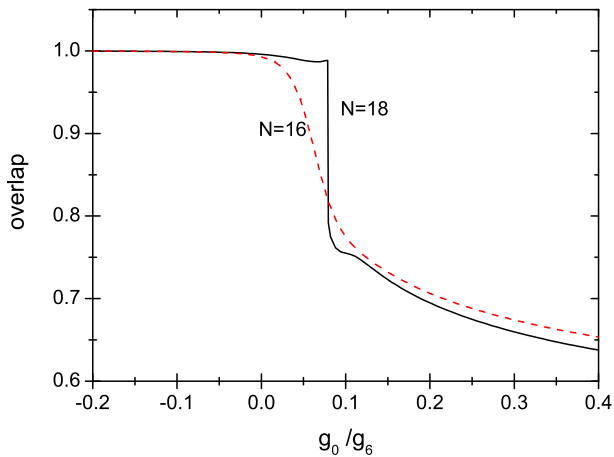}}
 \caption{(Color online) The overlap
$|\langle\tilde{\psi}_1|\psi_1\rangle|$ of the exact ground
state $\psi_1$ of $^{52}$Cr (the one expanded in Fock-states)
and the corresponding approximate state $\tilde{\psi}_1$ (using
the $S=0$ and 2 pairs as building blocks) against $g_0$. $N=18$
(solid) and 16 (dash), and $M=0$.}
 \label{fig4}
\end{figure}

The overlap $|\langle\tilde{\psi}_1|\psi_1\rangle|$ of the approximate and
exact ground states against $g_0$ is plotted in Fig.~\ref{fig4}. The solid
curve ($N=18$) is extremely close to one when $g_0$ is smaller than the
critical point. It confirms the physical picture that both the $S=0$ and 2
pairs are building blocks. However, the solid curve has a sudden fall at the
critical point. Thus the picture is spoiled when $g_0$ is larger, and we
have to look for other structures. The dash curve ($N=16$) represents the
case without a transition, where a swift descending replaces the sudden fall.

\section{Candidates of three-body substructures and the triplex-states}

In order to see whether 3-body substructures would exist in region III, let
us first analyze a 3-body spin-3 system. Let
\begin{equation}
 \xi_{\sigma\lambda m_{\lambda}}
  =  \beta_{\sigma\lambda}
     \tilde{P}_3
     ((\chi\chi)_{\sigma}\chi)_{\lambda m_{\lambda}},
 \label{e16_xi}
\end{equation}
be a spin-state of the 3-boson system, where two spins are
firstly coupled to $\sigma$. Then, they are coupled to
$\lambda$ and $m_{\lambda}$, the total spin and its
$Z$-component. $\beta_{\sigma\lambda}$ is for the
normalization. Totally there are 8 independent
$\xi_{\sigma\lambda}$ (the subscript $m_{\lambda}$ might be
neglected) given in the top row of Tab.~\ref{tab5}. Those not
listed in the table are linear combination of them (Say,
$\xi_{2,1}\equiv\xi_{4,1}$). Each of them represents a specific
3-body spin-structure. Incidentally, these eight
$\xi_{\sigma\lambda}$ are exactly orthogonal to each other
except $\xi_{2,3}$ and $\xi_{4,3}$.
\begin{table}[tbh]
 \caption{The probabilities $p_S^{\sigma\lambda}$ that a pair of
particles are coupled to $S$ in the 3-body states
$\xi_{\sigma\lambda}$.}
 \label{tab5}
\begin{ruledtabular}
 \begin{tabular}{lllllllll}
  $\sigma,\lambda$       &  $2,1$    &  $2,3$    &  $4,3$    &  $2,4$    &  $2,5$    &  $4,6$    &  $4,7$    &  $6,9$  \\
  \hline
  $p_0^{\sigma\lambda}$  &  0        &  $0.094$  &  $0.108$  &  0        &  0        &  0        &  0        &  0  \\
  $p_2^{\sigma\lambda}$  &  $0.524$  &  $0.484$  &  $0.135$  &  $0.611$  &  $0.413$  &  0        &  0        &  0  \\
  $p_4^{\sigma\lambda}$  &  $0.476$  &  $0.211$  &  $0.753$  &  $0.061$  &  $0.234$  &  $0.727$  &  $0.515$  &  0  \\
  $p_6^{\sigma\lambda}$  &  0        &  $0.211$  &  $0.003$  &  $0.328$  &  $0.353$  &  $0.273$  &  $0.485$  &  1  \\
 \end{tabular}
 \end{ruledtabular}
\end{table}

By extracting two particles from $\xi_{\sigma\lambda}$, one can
define and calculate the probabilities $p_S^{\sigma\lambda}$
that the two spins are coupled to $S$. They are given in
Tab.~\ref{tab5}. If a 3-body substructure is a basic
constituent, it must be very stable. Since
$p_6^{\sigma\lambda}$ of $\xi_{2,1}$ and $\xi_{4,3}$ are zero
or extremely small, the repulsion arising from $g_6$ can be
avoided in them. Therefore $\xi_{2,1}$ and $\xi_{4,3}$ will be
relatively more stable when $g_6$ is more positive than the
others. In a further competition between $\xi_{2,1}$ and
$\xi_{4,3}$, if $g_4$ is negative and $g_2$ is positive,
$\xi_{4,3}$ would be more stable due to having a large
$p_4^{\sigma\lambda}$. On the contrary, if $g_4$ is positive
and $g_2$ is negative, $\xi_{2,1}$ would be more stable.

To evaluate the importance of a substructure quantitatively,
from a $N$-body spin-state $\psi_i$, we define
$\Phi_{\sigma\lambda m_{\lambda}}^i
\equiv\langle\xi_{\sigma\lambda m_{\lambda}}|\psi_i\rangle$
which is a $(N-3)$-body spin-state. Then, for arbitrary three
particles in $\psi_i$, the probability that they form the
$\xi_{\sigma\lambda}$ substructure is $Q_{\sigma\lambda}^i
\equiv\sum_{m_{\lambda}}\langle\Phi_{\sigma\lambda
m_{\lambda}}^i|\Phi_{\sigma\lambda m_{\lambda}}^i\rangle$. Note
that $Q_{\sigma\lambda}^i$ depends on the total spin of
$\psi_i$, $F(i)$, but not on $M(i)$ because of the summation
over $m_{\lambda}$. If the particle correlation is removed, the
corresponding probability would be
$Q_{\sigma\lambda}^{\mathrm{free}}
\equiv\frac{2\lambda+1}{\sum_{\lambda'}(2\lambda'+1)}$ where
the summation over $\lambda'$ covers the eight states listed in
Tab.~\ref{tab5} (i.e., $\lambda'=1$, 3, 3, 4, 5, 6, 7, and 9,
where $\lambda'=3$ should be counted twice). Then, we define
the ratio $\mathfrak{\rho}_{\sigma\lambda}^i \equiv
Q_{\sigma\lambda}^i/Q_{\sigma\lambda}^{\mathrm{free}}$. If this
quantity is much larger than one, the substructure
$\xi_{\sigma\lambda}$ is very preferred.

As an example, we go out from $^{52}$Cr for a while and assume
that $g_6=1$, $g_4=-1$, $g_2=0.2$, and $g_0=-0.2$. Since $g_4$
has been chosen to be rather negative, $\xi_{4,3}$ is expected
to be important. To verify, the ratios denoted as
$\mathfrak{\rho}_{\sigma\lambda}^1(\mathrm{A})$ for the ground
state with $N=18$ are listed in Tab.~\ref{tab6}.
\begin{table}[tbh]
 \caption{$\mathfrak{\rho}_{\sigma\lambda}^1$ of the ground
state with $N=18$. The parameters of interaction associated
with the cases A and B are given in the text.}
 \label{tab6}
\begin{ruledtabular}
 \begin{tabular}{lllllllll}
  $\sigma,\lambda$                                 &  $2,1$    &  $2,3$    &  $4,3$    &  $2,4$    &  $2,5$    &  $4,6$    &  $4,7$    &  $6,9$  \\
  \hline
  $\mathfrak{\rho}_{\sigma\lambda}^1(\mathrm{A})$  &  $0.099$  &  $0.119$  &  $4.425$  &  $0.032$  &  $0.069$  &  $1.813$  &  $0.967$  &  $0.204$  \\
  $\mathfrak{\rho}_{\sigma\lambda}^1(\mathrm{B})$  &  $3.485$  &  $1.180$  &  $1.011$  &  $2.399$  &  $1.231$  &  $0.728$  &  $0.752$  &  $0.390$  \\
 \end{tabular}
 \end{ruledtabular}
\end{table}

It is shown that $\mathfrak{\rho}_{4,3}^1(\mathrm{A})$ is
particularly large. Therefore $\xi_{4,3}$ is very preferred by
the ground state as expected, and one might further expect that
the triplex $\xi_{4,3}$ might play a role as a building block.
To clarify, we introduce a set of basis functions for the case
with $N=3K$ as
\begin{equation}
 \Phi_j^{\mathrm{triplex}}
  =  \beta'_j
     \tilde{P}_N
     \{ [\xi_{4,3,3}]^{K_3}
        [\xi_{4,3,2}]^{K_2}
        \cdots
        [\xi_{4,3,-3}]^{K_{-3}} \},
 \label{e17_Phij}
\end{equation}
in which $\sum_{\mu}K_{\mu}=N/3$ and $\sum_{\mu}\mu K_{\mu}=M$.
For $N=18$ and $M=0$, there are totally 58 basis functions,
much smaller than 3486. After a diagonalization of
$H_{\mathrm{mod}}$ in the 58-dimensional space, we obtain the
approximate eigenstates $\psi_i^{\mathrm{triplex}}=\sum_j
d_j\Phi_j^{\mathrm{triplex}}$ to be compared with the exact
eigenstates $\psi_i$. It turns out that
$|\langle\psi_i^{\mathrm{triplex}}|\psi_i\rangle |=0.999$,
$0.996$, and $0.990$ for $i=1$, 2, and 3, respectively (they
are the three lowest states having $F(i)=0$, 4, and 6,
respectively). Such a great overlap confirms that, similar to
the spin-2 condensates, the triplex-structure exists also in
spin-3 condensates. However, the triplex of spin-2 condensates
has $\lambda=0$, thus there is only one kind of building
blocks. Whereas the triplex now has $\lambda=3$ and therefore
has $2\lambda+1=7$ kinds of building blocks. This leads to
complexity.

We go back to the case of $^{52}$Cr atoms. Let $g_2$, $g_4$,
and $g_6$ be given at the experimental values and $g_0/g_6$ is
given at some presumed values. Note that, in addition to $g_6$,
$g_4$ is also quite positive. Since $\xi_{4,3}$ has a large
$P_4^{\sigma\lambda}$, it is no more superior. Instead,
$\xi_{2,4}$ might be important due to having a very small
$P_4^{\sigma\lambda}$. When $g_0/g_6=0.5$, the ratios denoted
as $\mathfrak{\rho}_{\sigma\lambda}^1(\mathrm{B})$ has been
calculated and listed in the bottom row of Tab.~\ref{tab6}.
Where, both $\mathfrak{\rho}_{2,1}^i$ and
$\mathfrak{\rho}_{2,4}^i$ are large as expected. Therefore,
totally the twelve $\xi_{2,1,m_{\lambda}}$ and
$\xi_{2,4,m_{\lambda}}$ are used as building blocks, and we
define another set of basis functions as
\begin{eqnarray}
 \Phi_j^{\mathrm{tri},\mathrm{tri}}
 &=& \beta''_j
     \tilde{P}_N
     \{ ( \prod_{m_a}
          [\xi_{2,1,m_a}]^{K_{a,m_a}} )  \nonumber \\
  &&    ( \prod_{m_b}
          [\xi_{2,4,m_b}]^{K_{b,m_b}} ) \},
 \label{e18_Phij}
\end{eqnarray}
where $m_a$ is from $-1$ to 1, $m_b$ is from $-4$ to 4,
$\sum_{m_a}K_{a,m_a}+\sum_{m_b}K_{b,m_b}=N/3$, and
$\sum_{m_a}m_a K_{a,m_a}+\sum_{m_b}m_b K_{b,m_b}=M$. The number
of $\Phi_j^{\mathrm{tri},\mathrm{tri}}$ is 758 when $N=18$ and
$M=0$. However, only 615 of them are linearly independent. With
$\Phi_j^{\mathrm{tri},\mathrm{tri}}$, we have calculated the
approximate eigenstate $\psi_{i'}^{\mathrm{tri},\mathrm{tri}}$
at five values of $g_0/g_6$, and the overlaps
$|\langle\psi_{i'}^{\mathrm{tri},\mathrm{tri}}|\psi_i\rangle|$
are listed in Tab.~\ref{tab7}. It is reminded that
$g_0/g_6=0.079$ is a critical point. Once $g_0/g_6$ is larger
than the critical point, the overlaps are very close to one.
Thus the picture of triplexes is theoretically confirmed.
Whereas this picture is not well established when $g_0$ is
smaller than the critical point, where the $S=0$ and 2 pairs
are dominant.
\begin{table*}[tbh]
 \caption{The overlap
$|\langle\psi_{i'}^{\mathrm{tri},\mathrm{tri}}|\psi_i\rangle|$,
where $\psi_{i'}^{\mathrm{tri},\mathrm{tri}}$ is the
triplex-state formed by using $\xi_{2,1,m_{\lambda}}$ and
$\xi_{2,4,m_{\lambda}}$ as building blocks. $N=18$ and 12, and
$M=0$ are given. The parameters $g_2$, $g_4$, and $g_6$ are
from the experimental data of $^{52}$Cr, while $g_0/g_6$ is
denoted as $g'$ given at five values. $i'$ is so chosen that,
if $\psi_i$ is the $k$-th eigenstate of the series with
$F=F(i)$, then $\psi_{i'}^{\mathrm{tri},\mathrm{tri}}$ is also
the $k$-th state of the series of triplex-states with
$F(i')=F(i)$. $F(i)$ are given in the parentheses following the
overlaps.}
 \label{tab7}
\begin{ruledtabular}
 \begin{tabular}{llllll}
    & $g'=-1$ & $g'=-0.5$ & $g'=0.1$ & $g'=0.5$ & $g'=1$  \\
  \hline
  $|\langle\psi_{i'_1}^{\mathrm{tri},\mathrm{tri}}|\psi_1\rangle|_{N=18}$ &  $0.908$ (0)  &  $0.958$ (0)  &  $1.000$ (2)  &  $0.998$ (2)  &  $0.983$ (2)  \\
  $|\langle\psi_{i'_2}^{\mathrm{tri},\mathrm{tri}}|\psi_2\rangle|_{N=18}$ &  $0.859$ (2)  &  $0.935$ (2)  &  $1.000$ (0)  &  $0.998$ (2)  &  $0.998$ (4)  \\
  $|\langle\psi_{i'_3}^{\mathrm{tri},\mathrm{tri}}|\psi_3\rangle|_{N=18}$ &  $0.915$ (4)  &  $0.959$ (4)  &  $1.000$ (2)  &  $0.999$ (0)  &  $0.981$ (2)  \\
  $|\langle\psi_{i'_1}^{\mathrm{tri},\mathrm{tri}}|\psi_1\rangle|_{N=12}$ &  $0.861$ (0)  &  $0.934$ (0)  &  $1.000$ (0)  &  $0.997$ (2)  &  $0.988$ (2)  \\
  $|\langle\psi_{i'_2}^{\mathrm{tri},\mathrm{tri}}|\psi_2\rangle|_{N=12}$ &  $0.756$ (2)  &  $0.869$ (2)  &  $1.000$ (2)  &  $0.999$ (0)  &  $0.998$ (0)  \\
  $|\langle\psi_{i'_3}^{\mathrm{tri},\mathrm{tri}}|\psi_3\rangle|_{N=12}$ &  $0.813$ (4)  &  $0.905$ (4)  &  $1.000$ (3)  &  $0.999$ (1)  &  $0.998$ (1)  \\
 \end{tabular}
 \end{ruledtabular}
\end{table*}

\section{Final remarks}

Instead of using the MFT, a language from the few-body theory
is used in this paper. We have shown theoretically the
existence of stable 2-body and 3-body structures as building
blocks in small spin-3 condensates. The ratios
$\mathfrak{P}_S^i/\mathfrak{P}_S^{\mathrm{free}}$ and
$Q_{\sigma\lambda}^i/Q_{\sigma\lambda}^{\mathrm{free}}
\equiv\mathfrak{\rho}_{\sigma\lambda}^i$ defined in this paper
are crucial in the search of these basic constituents. The
reason leading to the appearance of these constituents is
explained based on the feature of interaction. Whereas, in the
MFT, the physics underlying the appearance of a specific spinor
is not easy to clarify.

The calculation in this paper concerns only small spin-3
condensates ($N\leq 24$). For spin-2 condensates, it has been
proved theoretically that the fact that pairs and triplexes
appear as building blocks does not depend on $N$ (In fact, the
picture of the triplexes would become even clearer when
$N\rightarrow\infty$ \cite{r_PVI}). It has also been proved
that the existence of the pairs in spin-3 condensates does not
depend on $N$.\cite{r_PVI,r_CGB2} Thus the existence of
triplexes as building blocks in large spin-3 condensates is
very probable, nonetheless it deserves a further study.

\begin{acknowledgments}
The support from the NSFC under the grant 10874249 is appreciated.
\end{acknowledgments}

\appendix

\section{Iteration relations}

For the pair-state and odd-pair-state the following equations
relating a $N$-body and a $(N-2)$-body systems are very useful:
\begin{eqnarray}
 \label{eA1}
 \mathcal{N}_0^{(N)}
 &\equiv&
     \langle
     \tilde{P}_N
     [(\chi\chi)_0]^{N/2} |
     \tilde{P}_N[(\chi\chi)_0]^{N/2}
     \rangle,  \nonumber \\
 &=& \frac{1}{7}
     N^2
     (N-1)
     (N+5)
     \mathcal{N}_0^{(N-2)}, \\
 \label{eA2}
 \mathcal{N}_{\mathrm{odd}}^{(N)}
 &\equiv&
     \langle
     \tilde{P}_N
     \chi_{\mu}
     [(\chi\chi)_0]^{(N-1)/2} |
     \tilde{P}_N
     \chi_{\mu}
     [(\chi\chi)_0]^{(N-1)/2}
     \rangle  \nonumber \\
 &=& \frac{1}{7}
     N
     (N-1)^2
     (N+6)
     \mathcal{N}_{\mathrm{odd}}^{(N-2)}, \\
 \label{eA3}
 \mathcal{N}_{S,m_S}^{(N)}
 &\equiv&
     \langle
     \tilde{P}_N
     (\chi\chi)_{S m_S}
     [(\chi\chi)_0]^{(N-2)/2} |
     \tilde{P}_N
     (\chi\chi)_{S m_S} \nonumber \\
  && [(\chi\chi)_0]^{(N-2)/2}
     \rangle  \nonumber \\
 &=& \frac{1}{7}
     N
     (N-1)
     (N-2)
     (N+7)
     \mathcal{N}_{S,m_S}^{(N-2)},
\end{eqnarray}
where $\tilde{P}_N$ denotes a summation over the $N!$
permutation terms, Eq.~(\ref{eA3}) holds only if $S\neq 0$ (if
$S=0$, then Eq.~(\ref{eA1}) should be used). Making use of
these equations related matrix elements can be derived via
iteration. For examples, the constant of normalization $\gamma$
can be obtained from Eq.~(\ref{eA1}), $\gamma'$ from
Eq.~(\ref{eA2}).

\section{The 1-body probabilities of odd-pair-states}

One can extract a particle (say, particle 1) from an odd-pair-state as
\begin{eqnarray}
 \tilde{P}_N
 \chi_{\mu}
 [(\chi\chi)_0]^K
 &=& \sum_{\nu}
     \chi_{\nu}(1)
     \{ \delta_{\mu,\nu}
        \tilde{P}_{N-1}
        [(\chi\chi)_0]^K  \nonumber \\
  &&   -(N-1)
        \frac{(-1)^{\nu}}{\sqrt{7}}
        \sum_S
        C_{3\mu,3,-\nu}^{S,\mu-\nu} \nonumber \\
  &&    \tilde{P}_{N-1}
        (\chi\chi)_{S,\mu-\nu}
        [(\chi\chi)_0]^{K-1} \},
 \label{eA4}
\end{eqnarray}
where $K=(N-1)/2$, the Clebsch-Gordan coefficients have been introduced, and
only even $S$ are included in the summation. Then, from the definition of
the 1-body probability, we have
\begin{eqnarray}
 P_{\nu}^{\mathrm{odd-polar}}
 &=& \delta_{\mu,\nu}
     \frac{4K+7}{7}
     \cdot
     \frac{\mathcal{N}_0^{(N-1)}}
          {\mathcal{N}_{\mathrm{odd}}^{(N)}}  \nonumber \\
  &&+\frac{4K^2}{7}
     \sum_S
     (C_{3\mu,3,-\nu}^{S,\mu -\nu})^2
     \frac{\mathcal{N}_{S,\mu-\nu}^{(N-1)}}
          {\mathcal{N}_{\mathrm{odd}}^{(N)}}.
 \label{eA5}
\end{eqnarray}

It is mentioned that the expression of $\mathcal{N}_{S,\mu-\nu}^{(N-1)}$
depends on whether $S$ is zero or nonzero. With this in mind, after a
simplification, Eq.~(\ref{eA5}) leads to the expressions given in the text.

\section{The correlative probabilities of pair-states}

One can extract two particles (say, 1 and 2) from a pair-state as
\begin{widetext}
\begin{eqnarray}
 \tilde{P}_N
 [(\chi\chi)_0]^{N/2}
 &=& N
     \sum_{\mu\nu}
     \chi_{\mu}(1)
     \chi_{\nu}(2)
     \{ \delta_{\mu,-\nu}
        C_{3\mu,3\nu}^{0,0}
        \tilde{P}_{N-2}
        [(\chi\chi)_0]^{(N-2)/2} \nonumber \\
  &&   +(N-2)
        \sum_S
        (2S+1)
        U
        \left(
        \begin{array}{ccc}
         3 & 3 & 0 \\
         3 & 3 & 0 \\
         S & S & 0
        \end{array}
        \right)
        C_{S,\mu+\nu,S,-\mu-\nu}^{0,0}
        C_{3\mu,3\nu}^{S,\mu+\nu}
        \tilde{P}_{N-2}
        (\chi\chi)_{S,-\mu-\nu}
        [(\chi\chi)_0]^{(N-4)/2} \},
 \label{eA6}
\end{eqnarray}
\end{widetext}where the Clebsch-Gordan and $9j$-coefficients have been
introduced, and only even $S$ are included in the summation.
Then, from the definition of the correlative probability, we
have
\begin{eqnarray}
 P_{\mu\nu}^{\mathrm{polar}}
 &=& (\frac{N}{7})^2
     [ \delta_{\mu,-\nu}
       (2N+3)
       \frac{\mathcal{N}_0^{(N-2)}}
            {\mathcal{N}_0^{(N)}}  \nonumber \\
  &&  +(N-2)^2
       \sum_S
       (C_{3\mu,3\nu}^{S,\mu+\nu})^2
       \frac{\mathcal{N}_{S,\mu+\nu}^{(N-2)}}
            {\mathcal{N}_0^{(N)}} ].
 \label{eA7}
\end{eqnarray}

After a simplification, Eq.~(\ref{eA7}) leads to the expressions given in
the text.


\begin{thebibliography}{99}
\bibitem{r_Z}
 The suppression of $V_{\mathrm{dd}}$ in strong isotropic
trap can also be seen under the frame work of MFT. In this
theory, the effect of $V_{\mathrm{dd}}$ is embodied by the
factor $c_d\Gamma_0$, where $c_d$ is the strength and is
usually small ($c_d=0.0036 g_6$ for $^{52}$Cr), and
$\Gamma_0=\int
\mathrm{d}\textbf{\textit{R}}\mathrm{d}\textbf{\textit{r}}
|\phi(\textbf{\textit{r}}_1)|^2|\phi(\textbf{\textit{r}}_2)|^2
Y_{20}(\hat{\textbf{\textit{r}}})/|\textbf{\textit{r}}|^3$,
where $\phi(\textbf{\textit{r}}_i)$ is the spatial wave
function of the $i$-th particle,
$\textbf{\textit{R}}=(\textbf{\textit{r}}_1+\textbf{\textit{r}}_2)/2$,
and
$\textbf{\textit{r}}=\textbf{\textit{r}}_1-\textbf{\textit{r}}_2$.
\cite{r_LSA} A part of the integrand can be in general expanded
as
$|\phi(\textbf{\textit{r}}_1)|^2|\phi(\textbf{\textit{r}}_2)|^2
=\sum_{l_a,l_b,L}B_{l_a,l_b,L}(R,r)(Y_{l_a}(\hat{\textbf{\textit{R}}})
Y_{l_b}(\hat{\textbf{\textit{r}}}))_L$ , where $l_a$ and $l_b$
denote the partial waves and they are coupled to $L$. Since
$\phi$ is close to be isotropic when the trap is strong, the
$L\neq 0$ terms can be neglected. This leads to $\Gamma_0=0$.
Thus the effect of $V_{\mathrm{dd}}$ is suppressed.
\end{thebibliography}
\end{document}